\documentclass[a4paper]{article}

\usepackage{INTERSPEECH2020}
\usepackage{multirow}

\title{VocGAN: A High-Fidelity Real-time Vocoder with a Hierarchically-nested Adversarial Network}
\name{Jinhyeok Yang$^1$, Junmo Lee$^1$, Youngik Kim$^1$, Hoonyoung Cho$^1$, Injung Kim$^2$}

\address{
    $^1$Speech AI Lab, NCSOFT, Republic of Korea\\
    $^2$School of CSEE, Handong Global University, Republic of Korea
}
\email{\{yangyangii,ljun4121,youngik,hycho\}@ncsoft.com, ijkim@handong.edu}

\begin{document}

\maketitle
\begin{abstract}

We present a novel high-fidelity real-time neural vocoder called VocGAN. A recently developed GAN-based vocoder, MelGAN, produces speech waveforms in real-time. However, it often produces a waveform that is insufficient in quality or inconsistent with acoustic characteristics of the input mel spectrogram. VocGAN is nearly as fast as MelGAN, but it significantly improves the quality and consistency of the output waveform. VocGAN applies a multi-scale waveform generator and a hierarchically-nested discriminator to learn multiple levels of acoustic properties in a balanced way. It also applies the joint conditional and unconditional objective, which has shown successful results in high-resolution image synthesis. In experiments, VocGAN synthesizes speech waveforms 416.7x faster on a GTX 1080Ti GPU and 3.24x faster on a CPU than real-time. Compared with MelGAN, it also exhibits significantly improved quality in multiple evaluation metrics including mean opinion score (MOS) with minimal additional overhead. Additionally, compared with Parallel WaveGAN, another recently developed high-fidelity vocoder, VocGAN is 6.98x faster on a CPU and exhibits higher MOS.

\end{abstract}

\section{Introduction}
\label{sec:intro}

Deep learning-based speech synthesis technology has rapidly improved in recent years. In particular, the emergence of neural vocoders, such as WaveNet \cite{oord2016wavenet} has significantly improved the fidelity of end-to-end speech synthesizers \cite{shen2018tacotron2}. However, for production-ready text-to-speech (TTS) systems, real-time generation on both GPU and CPU is also important. WaveNet is extremely slow because of the sequential auto-regressive structure. Parallel WaveNet \cite{oord2018parawavenet} and ClariNet \cite{ping2018clarinet} produce speech waveforms in real-time by leveraging a non-autoregressive parallel structure. However, they require a great deal of time and effort to reproduce performance because of the complex structure and the complicated training procedure (e.g., density distillation). On the other hand, flow-based models, such as WaveGlow \cite{prenger2019waveglow} and FloWaveNet \cite{kim2019flowavenet} can be directly learned by minimizing the negative log-likelihood of training data while generating high-quality speech waveforms at high speeds on a GPU. However, they require a huge number of parameters and heavy computation. Thus, it cannot synthesize speech waveforms in real-time without a GPU. 

Some recent models have adopted the idea of the generative adversarial network (GAN) to train neural vocoders. Parallel WaveGAN \cite{yamamoto2020parallel} applies a WaveNet-based generator conditioned on auxiliary features, such as mel spectrogram, and synthesizes natural-sounding speech waveforms in real-time on a GPU. The researchers of \cite{yamamoto2020parallel} trained the generator using adversarial loss and multi-resolution short-time Fourier transform (STFT) loss.
When combined with the Transformer TTS, Parallel WaveGAN exhibited a higher MOS than WaveNet and ClariNet. However, owing to its high computational complexity, Parallel WaveGAN could not achieve real-time on a CPU.

\begin{figure}[t]
  \centering
  \includegraphics[width=0.8\linewidth]{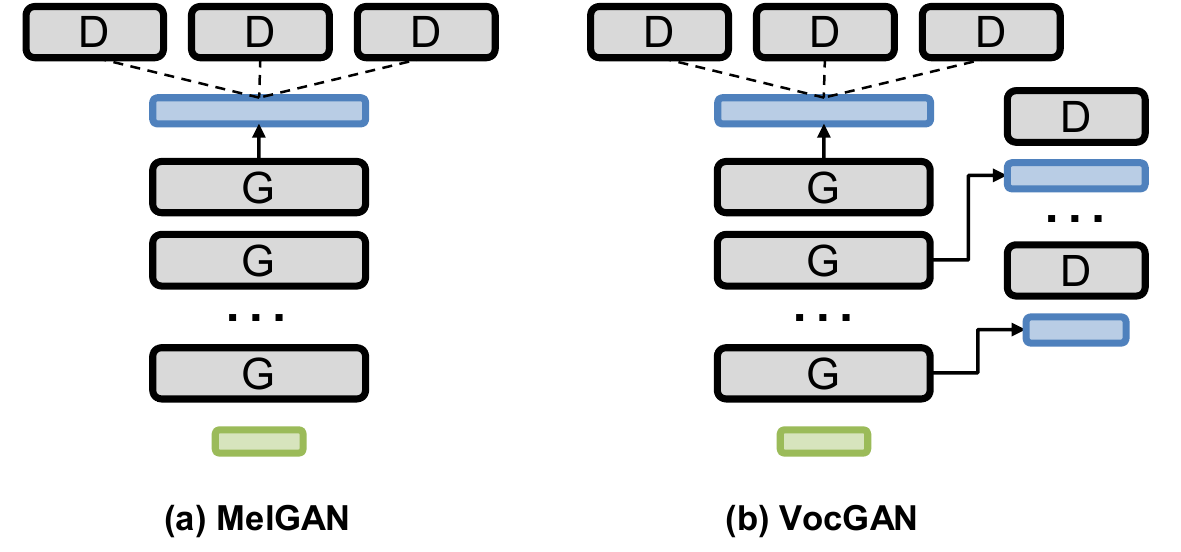}
  \caption{Comparison of MelGAN and VocGAN. The green boxes represent mel spectrograms, and the blue boxes represent output waveforms.}
  \label{fig:1}
\end{figure}

MelGAN \cite{kumar2019melgan} is another GAN-based vocoder that can be trained without density distillation or pretraining. In particular, MelGAN's generator consists of a carefully tuned lightweight network that enables real-time synthesis on a CPU.
MelGAN applies multiple techniques to produce raw waveforms of high temporal resolution. They apply the window-based objective \cite{isola2017img2img} to capture the high-frequency structure of an audio signal. Furthermore, they apply the multi-scale discriminator \cite{wang2018pix2pixhd} to learn the structure of audio at different levels. The kernel-size and stride of the transposed convolution are carefully chosen to avoid producing checkerboard artifacts.

In spite of the techniques, the output quality of MelGAN has room for improvement. It has been reported that applying instance normalization \cite{ulyanov2016instancenorm} washes away important pitch information, making the audio sound metallic. Although researchers have significantly alleviated this issue by replacing instance normalization with weight normalization \cite{salimans2016weightnorm}, the problem has not yet been completely solved.
The quality issues of MelGAN includes degradation of both low-frequency components (e.g.,fundamental frequency ($F_0$)) and high-frequency components (e.g.,noise). MelGAN also often produces a waveform inconsistent with acoustic characteristics of the input mel spectrogram. These issues suggest that the network structure and learning objective of MelGAN are not sufficient to correctly learn the acoustic representations of audio signals.

\begin{figure*}[t]
  \centering
  \includegraphics[width=0.82\linewidth]{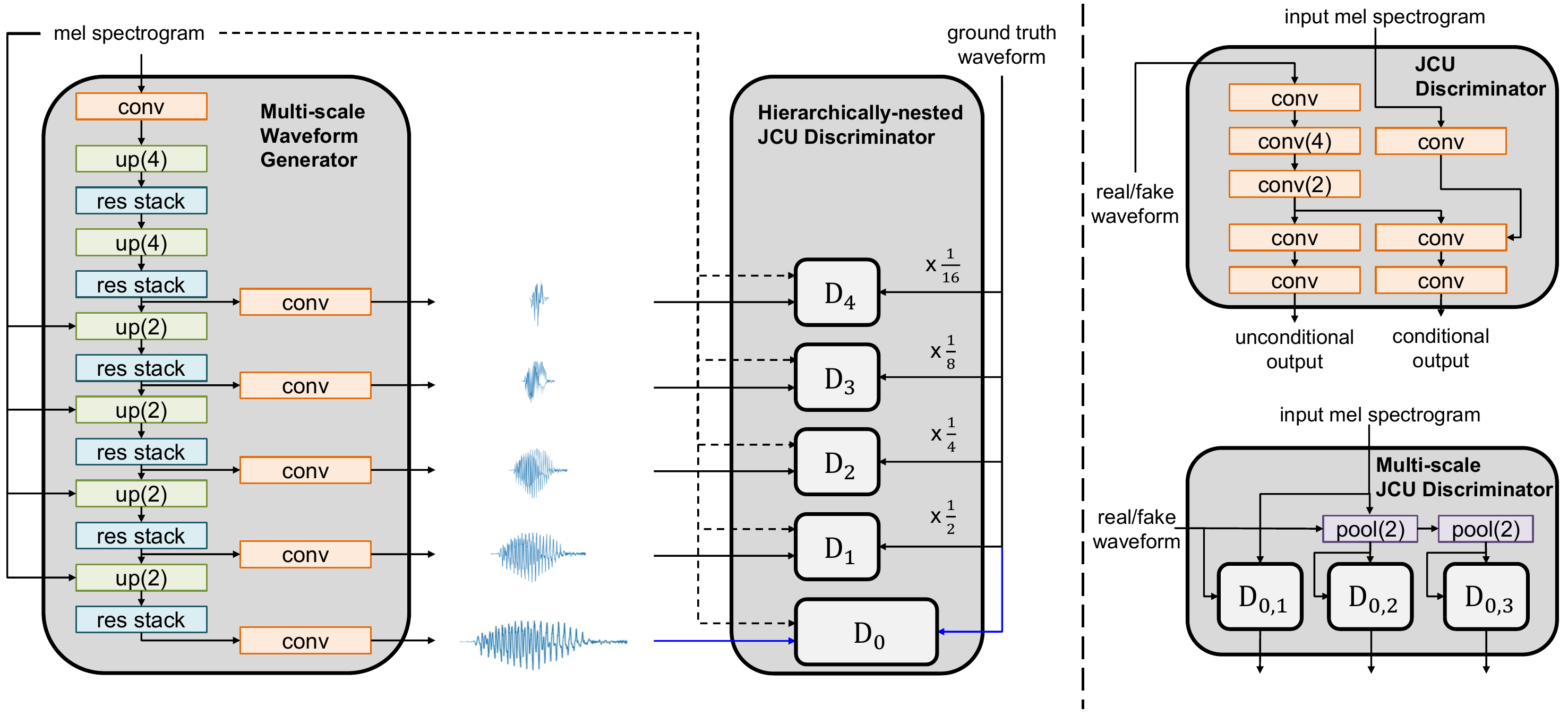}
  \caption{Model architecture. $\times \frac{1}{2^k}$ denotes a down-sampling rate. up($u$) denotes an up-sampling layer whose rate is $u$. conv and pool($v$) are convolutional layer and down-sampling pooling layer with a stride of $v$, respectively. res stack denotes a residual stack.}
  \label{fig:2}
\end{figure*}

In this paper, we present a novel GAN-based vocoder, VocGAN (Vocoder GAN), that is nearly as fast as MelGAN and significantly improves output quality and consistency with the input mel spectrogram.
VocGAN is an extension of MelGAN and applies multiple techniques to overcome its limitations. Fig. \ref{fig:1} illustrates the differences between MelGAN and VocGAN. Inspired by the work of \cite{zhang2018hdgan}, we extend the generator to output multiple waveforms at different scales. We train the generator with adversarial losses computed by a set of resolution-specific discriminators. This hierarchical structure helps to learn the various levels of acoustic properties in a balanced way. To each of the resolution-specific discriminators, we apply the joint conditional and unconditional (JCU) loss \cite{zhang2018stackgan++}, which has produced successful results in high-resolution image synthesis. We further improve the output quality by combining the STFT loss, which has exhibited good results in waveform synthesis \cite{yamamoto2020parallel}.

In experiments, VocGAN synthesized speech waveforms 416.7x faster on a GTX 1080Ti GPU and 3.24x faster on a CPU than real-time. Compared with MelGAN, it also exhibited significantly improved quality in mel cepstrum distance (MCD), $F_0$ root mean square error (RMSE), perceptual evaluation of speech quality (PESQ), and MOS with minimal additional overhead. Additionally, compared with with Parallel WaveGAN, VocGAN was 6.98x faster on a CPU and exhibits higher MOS.

\section{Proposed Method}
\label{sec:method}

\subsection{Baseline model}
\label{subsec:baseline}
To develop a high-fidelity neural vocoder that produces waveforms in real-time without a GPU, we choose MelGAN \cite{kumar2019melgan} as our baseline model. We improve MelGAN's model structure and its learning objective to overcome its quality degradation issues. The generator of the baseline model was implemented using a fully convolutional feed-forward network comprising four up-sampling blocks whose up-sampling rates are 8, 8, 2, and 2, respectively. Each up-sampling block contains a transposed convolution and a residual stack comprising three dilated convolution and a residual connection. The learning of the generator is guided using a multi-scale discriminator \cite{wang2018pix2pixhd} that computes an window-based objective \cite{isola2017img2img} from multiple waveforms down-sampled from the output waveform at different scales.

\subsection{Multi-scale waveform generator}
\label{sec:model}

The hierarchically-nested adversarial objective \cite{zhang2018hdgan} has shown effective in high-resolution image synthesis. It regularizes intermediate representations by inducing them to be useful for synthesizing waveforms at various resolutions. This helps the generator effectively learn not only the high-frequency components, but also the low frequency ones.
Although the hierarchically-nested adversarial objective was originally developed for image synthesis, its benefit may also be effective in improving the quality of raw waveforms as well. Therefore, we adopt the idea of \cite{zhang2018hdgan} to improve our neural vocoder.

Applying the hierarchically-nested objective to a neural vocoder requires major modifications to the generator and the discriminator. The generator takes as input a mel spectrogram and outputs the corresponding raw waveform. The design of our generator is based on the lightweight generator of MelGAN. However, we significantly modify its structure to apply the hierarchically-nested objective. As illustrated in the left part of Fig. \ref{fig:2}, the generator consists of six up-sampling blocks. The up-sampling rate of the first two up-sampling blocks is four, and that of the other blocks is two.
The generator outputs not only the final full-resolution waveform, but also multiple $k$ down-sampled waveforms ($1\leq k \leq K$) as side outputs, whose resolutions are $\frac{1}{2^k}$ of the full-resolution, respectively. $K$ denotes the number of down-sampled waveforms. In this research, $K$ is fixed at four. The $k$ down-sampled waveforms are generated from the output of the top-five up-sampling blocks via the convolution layers. The waveforms produced by the generator are formulated as Eq. (\ref{eq:1}).
\begin{equation}
    \label{eq:1}
    \hat{x}_0, ..., \hat{x}_K = G(s)
\end{equation}
where $s$ denotes the input mel spectrogram, $\hat{x}_0$ is the final full-resolution waveform, and $\hat{x}_1, ..., \hat{x}_K$ are the downsampled side waveforms.

Additionally, we add skip connections from the input mel spectrogram to each of the $2\times$ up-sampling blocks to learn intermediate representations directly conditioned on the input mel spectrogram. Thus, we improve acoustic characteristics consistency with the input mel spectrogram.

\begin{table*}[t]
  \caption{The result of ablation study. MCD(dB) and $F_0$ RMSE(Hz): the lower, the better. PESQ: the higher, the better.}
  \label{tab:1}
  \centering
  \begin{tabular}{l|ccc|ccc}
    \toprule
    \multirow{2}*{\textbf{Method}} &
    & \textbf{KSS} & & & \textbf{LJ} & \\
    &
    \textbf{MCD} &
    \textbf{$\boldsymbol{F_0}$ RMSE} &
    \textbf{PESQ} &
    \textbf{MCD} &
    \textbf{$\boldsymbol{F_0}$ RMSE} &
    \textbf{PESQ} \\
    \midrule
    Baseline (MelGAN)                                 &   4.478    &   38.80 & 2.51 & 4.614 & 50.04 & 2.74  \\
    \hline
    $+$ Hierarchically-nested structure and loss            &   3.986   &   37.84 & 2.66 & 3.827 & 49.39 & 2.91  \\
    $+$ JCU loss                                   &   3.441   &   35.39 & 2.93 & 3.551 & 45.87 & 3.06  \\
    $+$ Hierarchically-nested structure and loss $+$ JCU loss &   3.229   &   \textbf{32.36} & 3.37 & \textbf{3.144} & 44.19 & 3.32  \\
    $+$ Hierarchically-nested structure and loss $+$ STFT loss  &   3.438    &   34.99 & 3.03  & 3.707 & 48.68 & 3.03 \\
    \hline
    VocGAN    &   \textbf{2.974} &  32.85 & \textbf{3.48}  & 3.199 & \textbf{43.10} & \textbf{3.44}  \\
    \hline
    Ground Truth                                 &   0.0        &   0.0     & 4.5   & 0.0 & 0.0 & 4.5  \\
    \bottomrule
  \end{tabular}
\end{table*}

\subsection{Hierarchically-nested JCU discriminator}

\subsubsection{Hierarchically-nested structure}

The hierarchically-nested discriminator of VocGAN comprises five resolution-specific discriminators, as illustrated in the middle part of Fig. \ref{fig:2}. Each discriminator determines whether the output waveform of the corresponding resolution is real or fake. The hierarchically-nested discriminator leads the multi-scale waveform generator to learn the spectrogram-to-waveform mapping at five different resolutions. Thus, it helps the generator learn the mapping of both low- and high-frequency components of acoustic features. Each of the resolution-specific discriminators applies the JCU loss described in the next subsection. Moreover, for the final discriminator, we apply a multi-scale JCU discriminator whose structure is similar to the multi-scale discriminator of pix2pixHD \cite{wang2018pix2pixhd} and MelGAN \cite{kumar2019melgan}.

Our hierarchically-nested discriminator differs from the multi-scale discriminator of \cite{kumar2019melgan} and \cite{wang2018pix2pixhd} in the following ways. Ours takes as input multiple reduced-resolution samples that are directly generated from the intermediate representations of the generator, and the multi-scale discriminator takes a single full-resolution waveform and produces the reduced-resolution waveform by down-sampling the input waveform at various rates. The former forces multiple intermediate layers to learn how to generate reduced-resolution waveforms. As a result, it directly induces the generator to learn the high- and low-frequency features in more balanced way, which is important for speech signal processing.

We apply the least-square adversarial objective \cite{mao2017least} for the resolution-specific discriminators, as shown in Eq. (\ref{eq:2}).
\begin{equation}
    \setlength\abovedisplayskip{3pt}
    \setlength\belowdisplayskip{3pt}
    \label{eq:2}
    V_k(G,D_k) = \frac{1}{2}\mathbb{E}_s[D_k(\hat{x}_k)^2] + \frac{1}{2}\mathbb{E}_{x_k}[(D_k(x_k)-1)^2]
\end{equation}
where $x_k$ is the $k$-down-sampled ground-truth waveform, and $V_k(G,D_k)$ denotes the objective function for the $k$-down-sampled waveform. Because the final discriminator comprises a multi-scale discriminator, its objective $V_0(G,D_0)$ is defined by the sum of objectives of the sub-discriminators.
The loss functions of the discriminator and generator are defined as Eq. (\ref{eq:3}) and (\ref{eq:4}), respectively.
\begin{equation}
    \setlength\abovedisplayskip{0pt}
    \setlength\belowdisplayskip{0pt}
    \label{eq:3}
    L_{\mathrm{D}}(G,D) = \sum_{k=0}^K V_k(G,D_k)
\end{equation}
\begin{equation}
    \setlength\abovedisplayskip{0pt}
    \setlength\belowdisplayskip{0pt}
    \label{eq:4}
    L_{\mathrm{G}}(G, D) = \sum_{k=0}^K \frac{1}{2}\mathbb{E}_s[(D_k(\hat{x}_k)-1)^2]
\end{equation}

\subsubsection{Joint conditional and unconditional loss}
\label{subsec:jcu}

We combine the idea of JCU loss \cite{zhang2018stackgan++} to the hierarchically-nested adversarial objective to further improve speech quality. The right part of Fig. \ref{fig:2} shows the structures of our JCU discriminators. In contrast to conventional adversarial loss, JCU loss combines the conditional and unconditional adversarial losses as Eq. (\ref{eq:5})-(\ref{eq:7}). The conditional loss leads the generator to map the acoustic feature of the input mel spectrogram to the waveform more accurately. Thus, it helps reduce the discrepancy between the acoustic characteristics of the input mel spectrogram and the output waveform.
\begin{multline}
    \setlength\abovedisplayskip{0pt}
    \setlength\belowdisplayskip{0pt}
    \label{eq:5}
    V_k^{JCU}(G,D_k) = \frac{1}{2}\mathbb{E}_s[D_k(\hat{x}_k)^2 + D_k(\hat{x}_k,s)^2]\\
    + \frac{1}{2}\mathbb{E}_{(s,x_k)}[(D_k(x_k)-1)^2+(D_k(x_k,s)-1)^2]
\end{multline}
\begin{equation}
    \setlength\abovedisplayskip{0pt}
    \setlength\belowdisplayskip{0pt}
    \label{eq:6}
    L_{\mathrm{D}}^{JCU}(G,D) = \sum_{k=0}^K V_k^{JCU}(G,D_k)
\end{equation}
\begin{equation}
    \setlength\abovedisplayskip{0pt}
    \setlength\belowdisplayskip{0pt}
    \label{eq:7}
    L^{JCU}_{\mathrm{G}}(G, D) = \sum_{k=0}^K \frac{1}{2}\mathbb{E}_s[(D_k(\hat{x}_k)-1)^2+(D_k(\hat{x}_k,s)-1)^2]
\end{equation}

\subsubsection{Feature matching loss}
\label{subsec:fm}

Similar to MelGAN, we use the feature-matching loss that was first introduced in \cite{wang2018pix2pixhd}. This loss is defined by the $L_1$ distance between the discriminator feature maps computed from the real and synthesized waveforms as Eq. (\ref{eq:8}). We apply the feature matching loss to all of the resolution-specific discriminators. This loss significantly stabilizes the training.
\begin{equation}
    \label{eq:8}
    L_{\text{FM}}(G, D) = \mathbb{E}_{(s,x)}\bigg[\sum_{k=0}^K\sum_{t=1}^{T_k} \frac{1}{N_t}||D_k^{(t)}(x_k) - D_k^{(t)}(\hat{x}_k)||_1\bigg]
\end{equation}
$T_k$ is the total number of layers in the $k^{th}$ resolution-specific discriminator, and $N_t$ is the number of elements in each layer.

\subsection{Multi-resolution STFT loss}
\label{subsec:stft}

To improve the stability and efficiency of the adversarial training, we apply the multi-resolution STFT loss introduced in \cite{yamamoto2020parallel} as an auxiliary loss. In particular, this loss accelerates the speed of the training convergence. The auxiliary loss is used for the generator independently of the adversarial objectives. A single STFT loss measures the frame-level difference between the ground truth and the synthesized full-resolution waveform. The multi-resolution STFT loss $L_{STFT}$ is the sum of multiple STFT losses with different FFT sizes, window sizes, and frame shifts.

The total objective function for the generator combines all losses mentioned above. In this research, we set $\alpha=10.0$ and $\beta=1.0$, as shown in Eq. (\ref{eq:9}).
\begin{equation}
    \label{eq:9}
    L^{total}_G(G, D) =L^{JCU}_{G}(G, D) + \alpha L_{FM}(G,D) +\beta L_{STFT}(G)
\end{equation}

\section{Experiments}

\subsection{Datasets and experimental settings}

For experiments, we used the Korean Single Speaker Speech (KSS) dataset \cite{KSSdataset} and the LJSpeech dataset \cite{lj}. The KSS dataset contains 12,853 scripts and accompanying audio samples recorded by a single Korean female speaker. The LJSpeech dataset contains 13,100 samples recorded by a single American female speaker. The total lengths are 12 and 24 hours, respectively. We unified the sampling rates of the two datasets to 22,050 Hz for training. We used 129 and 131 utterances for validation and another 129 and 131 for testing. All remaining samples were used for training. We conducted our experiment on a server with an NVIDIA Tesla V100 GPU for training and an Intel Xeon(R) E5-2620 v4 2.10GHz CPU and an NVIDIA GTX 1080Ti GPU for testing.

\subsection{Training and evaluation}
We trained all models for 3,000 epochs. We used the Adam optimizer \cite{kingma2014adam} with a learning rate of 0.0001 with $\beta1 = 0.5$ and $\beta2 = 0.9$ for both generator and discriminator. We cut the samples into 1 seconds audio clips and used them for training. For multi-resolution STFT loss, we applied three STFT losses with frame sizes of 512, 1,024 and 2,048, window sizes of 240, 600 and 1,200 and frame shifts of 50, 120 and 240, respectively.

We evaluated the proposed methods using three objective metrics and one subjective metric. We used the MCD \cite{tamamori2017wavenetvocoder} and $F_0$ RMSE between the ground truth and the synthesized waveform to measure how accurately the vocoder converted the mel spectrogram into a waveform. To evaluate waveform quality, we measured PESQ \cite{pesq}, and we used MOS to evaluate subjective speech quality.

\subsection{Experimental results}
\subsubsection{Ablation study}

\begin{figure}[t]
  \centering
  \centerline{\includegraphics[width=3.5cm]{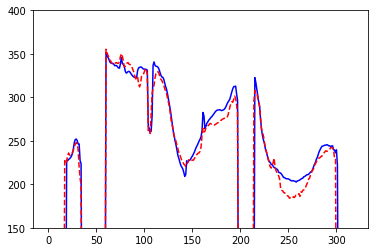}\includegraphics[width=3.5cm]{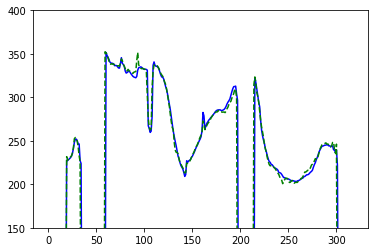}}
\caption{$F_0$ trajectories of speech waveforms. Blue solid (both): ground truth. Red dashed (left): MelGAN. Green dashed (right): VocGAN.}
\label{fig:3}
\end{figure}

We conducted an ablation study to analyze the effect of the proposed methods on both KSS and LJSpeech datasets. Starting from the baseline model, MelGAN, which applies the multi-scale discriminator and the feature-matching loss, we added each of the proposed methods one at a time measuring MCD, $F_0$ RMSE, and PESQ. Table \ref{tab:1} displays the results. Applying the hierarchically-nested objective and structures significantly improved the speech quality in all metrics, especially MCD. Replacing conventional least-squares loss with JCU loss drew significant improvement as well. The hierarchically-nested JCU objective that combines the two improvement techniques exhibited even more significant improvement. Combined with only the hierarchically-nested adversarial objective, multi-resolution STFT loss exhibited significant improvement in all metrics. However, it improved only PESQ slightly when combined with hierarchically-nested JCU objective. Additionally, we observed that the multi-resolution STFT loss accelerated the learning speed during the early stage. As shown in Fig. \ref{fig:3}, the $F_0$ trajectory of the synthesized waveform using the proposed method was significantly closer to the ground truth than that of the waveform synthesized by the baseline model.
\begin{table}[th]
  \caption{MOS with 95\% confidence intervals. The unit of inference speed is real-time factor that measures how many times faster than real-time.}
  \label{tab:2}
  \centering
  \begin{tabular}{cccc}
    \toprule
    \multirow{2}*{\textbf{Method}} & \multirow{2}*{\textbf{MOS}} & \multicolumn{2}{ c }{\textbf{Inference Speed}} \\
     & &  \textbf{GPU}  &  \textbf{CPU}  \\
    \midrule
    MelGAN                              &   3.898$\pm$0.091    &   \textbf{574.7x}   & \textbf{3.73x}   \\
    Parallel WaveGAN                    &   4.098$\pm$0.085    &   125.0x   & 0.47x       \\
    VocGAN (proposed)                    &   \textbf{4.202$\pm$0.081}    &   416.7x   & 3.24x     \\
    \hline
    Ground Truth                        &   4.721$\pm$0.052    &   -        & -     \\
    \bottomrule
  \end{tabular}
\end{table}
\subsubsection{Comparison with existing models}

We compared the proposed method with two recently developed neural vocoders: MelGAN and Parallel WaveGAN. To compare subjective speech quality, we measured the MOS score of the speech waveforms synthesized by each vocoder. First, we synthesized raw waveforms from 20 mel spectrograms randomly selected from the KSS test dataset using each vocoder\footnote{The audio samples are presented in the following URL:\\ \url{https://nc-ai.github.io/speech/publications/vocgan/}}. Then, 14 subjects rated the quality of the synthesized voices with a score in the range of 1 (worst) to 5 (best). Additionally, we measured the synthesis speed of the three vocoders. We used only one CPU core, and optionally, one GPU to measure synthesis speed.

The results are presented in Table \ref{tab:2}. VocGAN outperformed MelGAN and Parallel WaveGAN in MOS scoring. The Parallel WaveGAN exhibited a higher MOS than did MelGAN. However, the speed of Parallel WaveGAN was $4.6$x and $7.94$x slower than MelGAN on a GPU and on a CPU, respectively. The main reason of the slow speed is that the generator of Parallel WaveGAN is based on a WaveNet-based model that was significantly heavier than those of MelGAN and VocGAN. MelGAN was the fastest, and VocGAN was slower than MelGAN, but the difference was not very large. The synthesis speed of VocGAN was $416.7$x and $3.24$x faster than real-time on a GPU and a CPU, respectively.

\section{Conclusion}

In this paper, we proposed VocGAN, a GAN-based high-fidelity real-time vocoder. Comprising a multi-scale waveform generator and a hierarchically-nested JCU discriminator, VocGAN produced high-quality waveforms. We further improved its stability and efficiency of learning and the quality of the waveform by applying multi-resolution STFT loss as an auxiliary loss. The proposed method was nearly as fast as MelGAN, but it improved speech quality and consistency with the input mel spectrogram. VocGAN exhibited outperforming speech quality compared with Parallel WaveGAN and MelGAN. The synthesis speed of VocGAN is $416.7$x and $3.24$x faster than real-time on a GPU and a CPU, respectively. We expect this work, when combined with a recent feed forward acoustic model \cite{ren2019fastspeech}, will help build a real-time system with CPU.

\vfill\pagebreak

\clearpage

\bibliographystyle{IEEEtran}

\bibliography{mybib}

\end{document}